\documentclass[a4paper,dvips,12pt]{article}
\usepackage{epsfig,graphics,graphicx}
\begin{document}

\pagestyle{myheadings}
\markright{\it CO13-6}
\vskip.5in
\begin{center}

%
%
\vskip.4in {\Large\bf Radiative corrections to the Casimir effect for
the massive scalar field}
\vskip.3in
%
%
%
F.\ A.\ Barone,\footnote{Email: \tt fabricio@if.ufrj.br} 
R.\ M.\ Cavalcanti,\footnote{Email: \tt rmoritz@if.ufrj.br}
and
C.\ Farina\footnote{Email: \tt farina@if.ufrj.br}\\
Instituto de F\'{\i}sica, Universidade Federal do Rio de Janeiro\\
Caixa Postal 68528, 21941-972 Rio de Janeiro, RJ
%
%
%
\end{center}
%
\vskip.2in
\begin{abstract}

We compute the $O(\lambda)$ correction to the Casimir
energy for the massive $\lambda\phi^4$ model
confined between a pair of parallel plates. The calculations
are made with Dirichlet and Neumann boundary conditions.
The correction is shown to be sensitive to the boundary conditions,
except in the zero mass limit, in which case our results
agree with those found in the literature.

\end{abstract}
%
%
%
%

Since the publication of Casimir's paper in 1948 
\cite{Casimir} on the 
attraction of two neutral, parallel and  perfectly conducting 
plates (an effect that bears his name since then), a lot of work 
on this subject has been done \cite{Mostepanenko}.
However, the great majority of papers 
concerning the Casimir effect found in literature deals with 
non-interacting fields. Quite a few papers are 
devoted to radiative corrections to the Casimir effect
(in the context of QED, see \cite{Mostepanenko,QED};
in the context of the $\lambda\phi^4$ theory, see
\cite{KrechDietrichPRA92,phi4}). For 
free fields, the Casimir energy is given only by the zero-point 
energy of the field, properly regularized and renormalized. 
Hence, different boundary conditions leading to the 
same eigenfrequencies yield the same Casimir energy. In other 
words, at one-loop level the Casimir effect is sensitive to
the eigenfrequencies, but blind to the eigenmodes. An interesting
question is whether this peculiarity 
is maintained if radiative corrections are 
taken into account. In the case of a massless scalar field with 
a quartic self-interaction confined between a pair of parallel plates, 
it has been shown that the first order radiative correction to
the Casimir energy is the same for Dirichlet and Neumann boundary conditions
\cite{KrechDietrichPRA92}. 
However, it is not clear whether this is also true in the case
of a massive scalar field or is a peculiarity of the massless field.
Even for the massless scalar field, it is not obvious whether this property will persist 
at higher orders in the coupling constant.
The purpose of this work is to investigate the former question, namely,
what is the effect of the mass on the first order radiative correction to the Casimir
energy of a self-interacting scalar field
confined between a pair of parallel plates.

We shall consider in this paper the massive $\lambda\phi^4$
model in four-dimensional space-time.
It is defined by the Euclidean Lagrangian 
density\footnote{Conventions: $\hbar=c=1$, $x=(\tau,{\bf r},z)$.}
\begin {equation}
{\cal L}_{\rm E}={1\over 2}\,(\partial_{\mu}\phi)^{2}+{1\over 2}\,m^{2}\phi^{2}
+{\lambda\over 4!}\,\phi^{4}+{\cal L}_{\rm CT},
\end {equation}
where ${\cal L}_{\rm CT}$ contains the usual renormalization counterterms.
The system is confined between two parallel plates, located at
the planes $z=0$ and $z=a$, where the field is submitted to one
of the following boundary conditions:
\begin{itemize}
\item Dirichlet boundary conditions (DD):
\begin {equation}
\phi(z=0)=\phi(z=a)=0;
\end {equation}
\item Neumann boundary conditions (NN):
\begin {equation}
{\partial\phi\over\partial z}{\bigg |}_{z=0}
={\partial\phi\over\partial z}{\bigg |}_{z=a}=0.
\end {equation}
\end{itemize}

In the noninteracting case (i.e., $\lambda=0$), 
the Casimir energy per unit
of area is the same for both kinds of boundary 
conditions \cite{AmbjornWolframAP83}: 
\begin{equation}
\label{E0}
E_{\rm DD}^{(0)}=E_{\rm NN}^{(0)}=-\frac{m^2}{16\pi^2 a}
\sum_{n=1}^{\infty}\frac{K_2(2amn)}{n^2}.
\end{equation}
Using perturbation theory, one obtains for the $O(\lambda)$ correction 
to the above result the following expression:
\begin {equation}
\label {Einicial}
E^{(1)}=\int_{0}^{a}dz\left[{\lambda\over 8}\,G^{2}(x,x)
+{\delta m^{2}\over 2}\,G(x,x)+\delta\Lambda\right],
\end {equation}
where $G(x,x')$ is the Green's function of the free theory (i.e.,
with $\lambda=0$, but obeying the boundary conditions),
$\delta m^2$ is the radiatively induced shift in the mass parameter,
and $\delta\Lambda$ is the shift in the cosmological constant (i.e.,
the change in the vacuum energy which is due solely to the
interaction, and not to the confinement).

In the spectral representation, the Green's function in $(d+1)$-dimensional
space-time is given by
\begin {equation}
G(x,x')=\int {d\omega\over 2\pi}\int {d^{d-1}k\over(2\pi)^{d-1}}
e^{-i\omega(\tau-\tau')+i{\bf k}\cdot({\bf r}-{\bf r}')}\,
\sum_{n}{\varphi_{n}(z)\,\varphi_{n}^{*}(z')\over
\omega^{2}+k^{2}+m^{2}+(n\pi/a)^{2}},
\end {equation}
with
\begin {eqnarray}
\varphi_{n}^{\rm DD}(z)&=&\sqrt{2\over a}\,\sin\left(\frac{n\pi z}{a}\right),
\quad n=1,2,3,\ldots,
\\
\varphi_{n}^{\rm NN}(z)&=&\sqrt{2-\delta_{n,0}\over a}\,
\cos\left(\frac{n\pi z}{a}\right),
\quad n=0,1,2,3,\ldots.
\end {eqnarray}
$G(x,x')$ diverges when $x'\to x$ for $d\ge 1$, therefore a regularization
prescription is needed in order to make sense of $G(x,x)$. We shall
compute it using dimensional regularization; the result is
\begin {equation}
\label {zxc1}
G(x,x)={\Gamma\left(1-d/2\right)\over(4\pi)^{d/2}}
\sum_{n}\omega_{n}^{d-2}\,\varphi_{n}(z)\,\varphi_{n}^{*}(z)\qquad(d<1),
\end {equation}
where
\begin {equation}
\omega_{n}=\sqrt{m^{2}+\left(\frac{n\pi}{a}\right)^{2}}.
\end {equation}
Now we can compute the terms appearing in Eq.\ (\ref{Einicial}):
\begin{equation}
\int_{0}^{a}dz\,G(x,x)=
{\Gamma\left(1-d/2\right)\over(4\pi)^{d/2}}
\sum_{n=n_0}^{\infty}\omega_{n}^{d-2},
\end{equation}
\begin {equation}
\int_{0}^{a}dz\, G^{2}(x,x)=
{\Gamma^{2}\left(1-d/2\right)\over 4\pi a}
\left[\left(\sum_{n=n_0}^{\infty}\omega_{n}^{d-2}\right)^{2}
+{1\over 2}\sum_{n=1}^{\infty}\omega_{n}^{2d-4}\right],
\end{equation}
where $n_0=1$ in the DD case, and $n_0=0$ in the NN case.

Collecting terms, we obtain
\begin {eqnarray}
E_{\rm DD}^{(1)}&=&{\lambda\over 8a}\left[{\Gamma\left(1-d/2\right)\over(4\pi)^{d/2}}
F(2-d)+{2a\,\delta m^{2}\over\lambda}\right]^{2}
\nonumber \\
& &+{\lambda\over 16a}\,{\Gamma^{2}\left(1-d/2\right)\over(4\pi)^{d}}
F(4-2d)+\left[{\delta\Lambda}-{(\delta m^{2})^{2}\over 2\lambda}\right]a,
\label{E2passo}
\\
E_{\rm NN}^{(1)}&=&{\lambda\over 8a}\left[{\Gamma\left(1-d/2\right)\over(4\pi)^{d/2}}
\left[F(2-d)+m^{d-2}\right]+{2a\,\delta m^{2}\over\lambda}\right]^{2}
\nonumber \\
& &+{\lambda\over 16a}\,{\Gamma^{2}\left(1-d/2\right)\over(4\pi)^{d}}
F(4-2d)+\left[\delta\Lambda-{(\delta m^{2})^{2}\over 2\lambda}\right]a,
\label{E3passo}
\end {eqnarray}
where $F(s)$ is defined as
\begin {equation}
F(s)\equiv\sum_{n=1}^{\infty}\omega_{n}^{-s}
=\sum_{n=1}^{\infty}\left[m^{2}+\left({n\pi\over a}\right)^{2}\right]^{-s/2},
\qquad\Re(s)>1.
\end {equation}
The identity \cite{AmbjornWolframAP83}
\begin {equation}
\label{extF}
F(s)=-{1\over 2}m^{-s}
+{am^{1-s}\over 2\pi^{1/2}\,\Gamma\left(s/2\right)}
\left[\Gamma\left({s-1\over 2}\right)+4\sum_{n=1}^{\infty}
{K_{(1-s)/2}(2amn)\over(amn)^{(1-s)/2}}\right],
\end {equation}
where $K_{\nu}$ denotes the modified Bessel function, provides
the analytic extension of $F(s)$ to $\Re(s)\le 1$,
with simple poles at $s=1,-1,-3,-5,\ldots$

Let us now fix the renormalization conditions for 
$\delta m^2$ and $\delta\Lambda$.
To do so, let us consider the self-energy $\Sigma$. In first order 
in perturbation theory it is given by
\begin {equation}
\label {defbulk}
\Sigma(x)={\lambda\over 2}G(x,x)+\delta m^{2}.
\end {equation}
$\delta m^{2}$ will be fixed by imposing the following conditions 
on $\Sigma(x)$: (i) that $\Sigma(x)<\infty$ (except possibly
at some special points), and (ii) that $\Sigma(x)$ vanishes
away from the plates when $a\to\infty$, i.e.,
\begin {equation}
\label {condbulk}
\lim_{a\to\infty}\Sigma(z=\gamma a)=0,\qquad 0<\gamma<1;
\end {equation}
besides, we shall require that $\delta m^2$ be independent
of $a$. These conditions are fulfilled by taking
$\delta m^2=-(\lambda/2)\,G_0(0)$, where $G_{0}(z)$ denotes the free Green's 
function {\em without boundary conditions} evaluated at the point $x=(0,{\bf 0},z)$.
(This is most easily seen in the multiple reflection representation
of the Green's function $G$, in which Eq.\ (\ref{defbulk}) becomes
\begin {equation}
\Sigma(x)={\lambda\over 2}\sum_{n=-\infty}^{\infty}
\left[G_{0}(2na)\pm G_0(2z+2na)\right]+\delta m^{2},
\end {equation}
where the $+(-)$ sign corresponds to the NN(DD) case.) Computing
$G_0(0)$ within dimensional regularization, we explicitly
obtain
\begin {equation}
\label{dm2}
\delta m^2=-\frac{\lambda}{2}\int {d\omega\over 2\pi}\int {d^{d}k\over(2\pi)^{d}}\,
{1\over\omega^{2}+k^{2}+m^{2}}
=-{\lambda\over 2}{\Gamma\left((1-d)/2\right)\over (4\pi)^{(d+1)/2}}\,m^{d-1}.
\end {equation}
For the shift in the cosmological constant we shall take
\begin {equation}
\label{dL}
\delta\Lambda={(\delta m^{2})^{2}\over 2\lambda}.
\end {equation}
With this choice one eliminates the terms proportional to $a$ in 
Eqs.\ (\ref{E2passo}) and (\ref{E3passo}), which do not contribute
to the force between the plates (the linear dependence on $a$ is canceled
by similar terms when one adds the energy of the regions $z<0$
and $z>a$).

Inserting Eqs.\ (\ref{extF}), (\ref{dm2}), and (\ref{dL}) 
into Eqs.\ (\ref{E2passo}) and (\ref{E3passo}), we obtain
\begin {eqnarray}
\label {E1dDD}
E_{\rm DD}^{(1)}&=&{\lambda\over 8a}\left[{4am^{d-1}\over(4\pi)^{(d+1)/2}}
\sum_{n=1}^{\infty}{K_{(d-1)/2}(2amn)\over(amn)^{(d-1)/2}}
-{\Gamma\left(1-d/2\right)\over 2(4\pi)^{d/2}}m^{d-2}\right]^{2}
\nonumber \\
& &+{\lambda\over 32a}{\Gamma^{2}\left(1-d/2\right)\over(4\pi)^{d}}
\left\{-m^{2d-4}+{am^{2d-3}\over\pi^{1/2}\Gamma(2-d)}\left[
\Gamma\left(\frac{3}{2}-d\right)\right. \right.
\nonumber \\
& &+\left. \left. 4\sum_{n=1}^{\infty}
{K_{(2d-3)/2}(2amn)\over(amn)^{(2d-3)/2}}\right]\right\},
\end {eqnarray}
\begin {eqnarray}
\label {E1dNN}
E_{\rm NN}^{(1)}&=&{\lambda\over 8a}\left[{4am^{d-1}\over(4\pi)^{(d+1)/2}}
\sum_{n=1}^{\infty}{K_{(d-1)/2}(2amn)\over(amn)^{(d-1)/2}}
+{\Gamma\left(1-d/2\right)\over 2(4\pi)^{d/2}}m^{d-2}\right]^{2}
\nonumber \\
& &+{\lambda\over 32a}{\Gamma^{2}\left(1-d/2\right)\over(4\pi)^{d}}
\left\{-m^{2d-4}+{am^{2d-3}\over\pi^{1/2}\Gamma(2-d)}\left[
\Gamma\left(\frac{3}{2}-d\right)\right. \right.
\nonumber \\
& &+ \left. \left. 4\sum_{n=1}^{\infty}
{K_{(2d-3)/2}(2amn)\over(amn)^{(2d-3)/2}}\right]\right\}.
\end {eqnarray}
We can now take $d=3$, obtaining
\begin {equation}
\label {E1d=3DD}
E_{\rm DD}^{(1)}={\lambda m^{2}\over 512 \pi^{2}a}\left[\left(1+{2\over\pi}
\sum_{n=1}^{\infty}{K_{1}(2amn)\over n}\right)^{2}-1\right],
\end {equation}
\begin {equation}
\label {E1d=3NN}
E_{\rm NN}^{(1)}={\lambda m^{2}\over 512 \pi^{2}a}
\left[\left(1-{2\over\pi}\sum_{n=1}^{\infty}{K_{1}(2amn)\over n}\right)^{2}-1\right].
\end {equation}

It follows froms Eqs.\ (\ref{E1d=3DD}) and (\ref{E1d=3NN})
that $E_{\rm DD}^{(1)}\ge E_{\rm NN}^{(1)}$ in three-dimensional space;
equality is attained only in the massless case, in which case one
recovers the results of Ref.\ \cite{KrechDietrichPRA92}:\footnote{In order 
to take the limit $m\to 0$ in Eqs.\ (\ref{E1d=3DD}) and (\ref{E1d=3NN}),
we have used the following expansion, valid for small $z$ \cite{Braden}:
$$
\sum_{n=1}^{\infty}{K_{1}(nz)\over n}={\pi^{2}\over 6z}
-{\pi\over 2}+O(z\,\ln z).
$$
}
\begin {equation}
\label {EDD}
E_{\rm DD}^{(1)}=E_{\rm NN}^{(1)}={\lambda\over 18432 a^{3}}
\qquad(d=3, m=0).
\end{equation}
It is also interesting to note that, while $E_{\rm DD}^{(1)}$ is a
monotonically decreasing function of $a$, $E_{\rm NN}^{(1)}$ first
decreases and then eventually increases with $a$. Indeed, 
Eq.\ (\ref{EDD}) is also valid for fixed $m$ and $a\to 0$; on the other hand,
for $a\gg m^{-1}$ Eqs.\ (\ref{E1d=3DD}) and (\ref{E1d=3NN})
yield
\begin{equation}
E_{\rm NN}^{(1)}\approx -E_{\rm DD}^{(1)}
\approx -\frac{\lambda m^{3/2}}{256\pi^{5/2}}\,a^{-3/2}\,\exp(-2ma).
\end{equation}

Results for other kinds of boundary conditions (DN,
periodic, and anti-periodic) will be presented elsewhere \cite{nostres}.

\section*{Acknowledgments}
 
This work was supported by CNPq and CAPES.

%
%
%
%
%
%

%
\end{document}